
%
%
\input harvmac
\noblackbox
\sequentialequations

\def\ajou#1&#2,#3(#4){\ #1{\bf#2},#3\rm(19#4)}
\font\ticp=cmcsc10
\def\dee{\partial}
\def\d{\delta}
\def\p{\prime}
\def\s{\sigma}
\def\z{\zeta}
\def\ginf{\rightarrow\infty}
\def\D{\Delta}
\def\wp{w(+)}
\def\wn{w(-)}
\def\t{\tau}
\def\dt{\delta t}
\def\Nmin{N_{\rm min}}
\def\nb{\bar n_2}
%
%
\lref\Gunton{J. D. Gunton, M. San Miguel, and P. S. Sahni, in
{\it Phase Transitions and Critical Phenomena}, edited by
C. Domb and J. L. Lebowitz (Academic, New York, 1983), Vol. 8,
p. 267\semi
K. Binder, Physica {\bf 140A}, 35 (1986)\semi
H. Furukawa, Adv. Phys. {\bf 34}, 703 (1985).}
\lref\Amit{see for example D. J. Amit, {\it Field Theory, the
Renormalization Group, and Critical Phenomena}, (World Scientific,
Singapore, 1984) p. 189 ff.}
\lref\BrayIsing{A.~J.~Bray,\ajou J.~Phys.~A &22, L67 (89).}
\lref\BrayRut{A.~J.~Bray and A.~D.~Rutenberg, unpublished.}
\lref\Cardy{J.~L.~Cardy,\ajou J.~Phys.~A &14, 1407 (81).}
\lref\LCA{I.~M.~Lifshitz,\ajou Sov.~Phys.~JETP &15, 939 (62)\semi
S.~M.~Allen and J.~W.~Cahn,\ajou Acta Metall. &27, 1085 (79).}
\lref\BrayRG{A.~J.~Bray,\ajou Phys.~Rev.~Lett. &62, 2841 (89)\semi
A.~J.~Bray,\ajou Phys.~Rev.~B &41, 6724 (90).}
\lref\Glauber{R.~J.~Glauber,\ajou J.~Math.~Phys. &4, 191 (63).}
\lref\PS{M.~K.~Phani, {\it et al.}, \ajou Phys.~Rev.~Lett. & 45, 366 (80)
\semi P.~S.~Sahni, {\it et al.}, \ajou Phys.~Rev.~B & 24, 410 (81).}
\lref\HH{P.~C.~Hohenberg and B.~I.~Halperin, Rev.~Mod.~Phys. {\bf 49}, 435
(1977).}
\lref\KW{K.~G.~Wilson and J.~Kogut, Phys.~Reports {\bf 12}, 75 (1974).}
\lref\Dyson{F.~J.~Dyson, Comm.~Math.~Phys. {\bf 12}, 91, 212 (1969).}
\lref\Racz{H.~Hayakawa, {\it et al.}, \ajou Phys.~Rev.~E & 47, 1499 (93)}
\footline={\hfill}
\Title{\vbox{\baselineskip15pt\hbox{UCSBTH-93-15}
}}
{\vbox{\centerline {Phase Ordering in One-Dimensional Systems}\vskip2pt
\centerline{with Long-Ranged Interactions}
}}
\centerline{{\ticp Benjamin P. Lee}\footnote{$^*$}
{email: blee@sarek.physics.ucsb.edu}$^\dagger$
{\ticp and John L. Cardy}\footnote{$^\dagger$}{Address after July 1993:
Department of Physics; Theoretical Physics; 1 Keble Road; Oxford OX1 3NP;
England.}}
\vskip.1in
\centerline{\sl Department of Physics}
\centerline{\sl University of California}
\centerline{\sl Santa Barbara, CA 93106-9530}
\bigskip
\centerline{\bf Abstract}
We study the dynamics of phase ordering of a non-conserved, scalar order
parameter in one dimension, with long-ranged interactions characterized by
a power law $r^{-d-\s}$.  In contrast to higher dimensional systems, the
point nature of the defects allows simpler analytic and numerical methods.
We find that, at least for $\s>1$, the model exhibits evolution to a
self-similar state characterized by a length scale which grows with time as
$t^{1/(1+\s)}$, and that the late time dynamics is independent of
the initial length scale.  The insensitivity of the dynamics to the initial
conditions is consistent with the scenario of an attractive, non-trivial
renormalization group fixed point which governs the late time behavior.  For
$\s\le 1$ we find indications in both the simulations and an analytic method
that this behavior may be system size dependent.

\Date{5/93}

\newsec{Introduction}

\footline={\hss\tenrm\folio\hss}

The problem of the dynamical behavior of systems quenched from a disordered
phase to an ordered
phase has proven to be difficult to solve \Gunton.  At late times the dynamics
is described by the motion of domain walls separating equilibrated
regions.   It is believed that the domain wall morphology is
self-similar in time, and characterized by a length scale growing
with a power law $t^\rho$.  The universal nature of the dynamics has
prompted attempts to incorporate a renormalization group scheme
in describing the asymptotically late times \BrayRG, and also to determine
the various universality classes.  The value of $\rho$ appears to
depend strongly on the presence or absence of conservation laws,
and also on the dimension of the order parameter.
Unlike equilibrium critical phenomena, however, the
dimension of the system does not play a key role.
The generally accepted value for a scalar, non-conserved order parameter
(to which we restrict ourselves in this paper)
with short-ranged interactions is $\rho=1/2$ \refs{\LCA{,}\PS}.

Recent interest has been directed toward including long-ranged interactions
in the systems, characterized by a power law $V(r)\sim -r^{-d-\s}$
\refs{\BrayRut{,}\Racz}.
Using energy dissipation arguments, Bray and Rutenberg have found the
growth law exponent to be modified by the long-ranged interactions \BrayRut.
In particular, they argue that, in the case of the scalar, non-conserved
order parameter system, $\rho=1/(1+\s)$ for $\s<1$, and that $\rho=1/2$
with logarithmic corrections for $\s=1$.  For $\s>1$, and $d>1$, they
recover the exponent for short-ranged interacting systems, $\rho=1/2$.

The addition of long-ranged interactions brings both complication
and simplification to the problem.  The
complication stems from the fact that the local evolution depends
on the global state of the system, which makes numerical simulation
particularly difficult.  However, one simplification is that the presence of
long-ranged interactions allows the possibility of studying
one-dimensional systems, since it is known that for $0<\s\le 1$ the
$d=1$ Ising model has a non-trivial phase transition and thus
an equilibrium two-phase region.
Also, for higher dimensional systems long-ranged forces may
dominate the curvature forces, allowing the latter to be neglected.
In this paper, however, we restrict ourselves to one dimensional systems.

The contents of the paper are as follows:
in the next section we discuss the model for our system, a
continuum Langevin equation without the noise term.  We show the
equivalence of this model to that of the dynamic Ising model
for non-conserved order parameter (Glauber dynamics \Glauber).  In section 3
we develop an approach based on renormalization group concepts, and propose
a general feature of the dynamics:  asymptotic lack of dependence on the
initial length scale.  We also present numerical data for $\s=1/2$ which
is in disagreement with the predicted value of $\rho=1/(1+\s)$, and also
exhibits dependence on the system size, $L$.  The lack of dependence on the
initial length scale and the system size dependence are the principal
results of this paper.  In section 4 we describe a fugacity
expansion which leads to direct computation of a Callan-Symanzik type $\beta$
function in the large $\s$ limit.  When we consider values of
$\s\le 1$ we find divergent terms appearing in our expansion, which
may be related to the anomalous behavior in the simulations.  However, in
contrast, the $Q\ginf$ Potts model has similar divergences, but when simulated
seems to give the expected value of $\rho=1/(1+\s)$.  In section 5 we
present an method for coupling the dynamics of the density and the two-particle
distribution function which leads to qualitatively accurate results.
In section 6 we discuss our simulation methods, and in
section 7 we present our conclusions.

\newsec{The Model}

In the following section we present a low-temperature mapping from the
long-ranged Ising Hamiltonian with spin degrees of freedom to a Hamiltonian
with domain wall degrees of freedom.  Next we introduce dynamics to the
system via Langevin equations without a noise term.  This is shown to be
equivalent to
Glauber dynamics when $\s\le 1$.  Finally, we discuss related models
which are motivated by the simplifications they offer.

\subsec{Ising Hamiltonian}

We begin by considering the one-dimensional Ising Hamiltonian
\eqn\isingham{H = -J\sum_{i < j}s_is_jV(x_i-x_j)}
where
\eqn\zaa{V(x_i-x_j)=|x_i-x_j|^{-(1+\s)}.}
and the lattice spacing $a=1$.  It is known \Dyson\ that
this system has a phase transition with some non-zero $T_c$ when $0<\s\le 1$,
and so there is a two-phase equilibrium region for $T<T_c$.
Since we are interested in the dynamics of the domain walls,
which are points objects in this one-dimensional case, it is convenient to map
this Hamiltonian with spin degrees of freedom to one with domain wall
degrees of freedom via a lattice equivalent of integration
by parts.  The resultant Hamiltonian is (apart from surface terms)
\eqn\zab{H=J\sum\limits_{i< j}s_i^\p s_j^\p U(x_i-x_j)}
where the lattice derivatives are defined as $s_i^\p=s_{i+1}-s_i$, and
the function $U(x_i-x_j)$, the lattice equivalent of the second
anti-derivative of $V(x_i-x_j)$, is defined by
\eqn\defu{V(r)=U(r+1)-2U(r)+U(r-1).}
The boundary conditions are chosen so that $U(r)$ contains no constant
or linear pieces, with the solution for $r\gg 1$
\eqn\zac{U(r)=\cases{{\displaystyle|r|^{1-\s}\over\displaystyle\s(1-\s)}+O(1/r)
&$\s\ne 1$\cr\strut&\cr -\log |r|+O(1/r)&$\s=1$.\cr}}
Since the limit of zero lattice spacing is well-behaved, and the important
contributions from the long-ranged interactions should be arising at
large $r$, the late time dynamics of the theory should be unaffected by
taking the continuum limit.

The $s_i^\p$ are zero everywhere neighboring spins are aligned, and equal to
$\pm 2$ at the domain boundaries.  Therefore the sum over spins can be
replaced by a sum over the positions of the domain walls.  The sign, or
charge, of
the domain walls will be alternating, with the consequence that nearest
neighbors will attract, next-nearest neighbors will repel, and so on.
Absorbing the coupling constant $J$ into a rescaling of the spins, we get the
Hamiltonian
\eqn\zad{H=\sum\limits_{i< j}(-1)^{i+j}U(x_i-x_j).}

\subsec{Dynamical Model}

To add dynamics to this Hamiltonian we use Langevin type equations of
motion, introducing a kinetic coefficient $\Gamma$.
\eqn\zae{{dx_i\over dt}=-\Gamma\,{\dee H\over\dee x_i}.}
We neglect any possible noise term, for reasons which we explain
below.  There is an additional rule to the dynamics.  When two charges meet
each other they annihilate, and are both removed from the system.  In the
original spin picture this corresponds to an island of up spins shrinking
to zero in a background of down spins, or vice versa.

These equations of motion for the domain walls are equivalent, for $\s\le 1$
and low temperatures, to using Glauber dynamics for the spins \Glauber.
To see this, consider a
Glauber dynamical Ising model with temperature $\beta$, lattice spacing $a$,
and characteristic free spin flip rate $\alpha$.  The flip rates for
interacting spins are found via detailed balance:
\eqn\zaf{{\wp\over \wn} = \exp(-\beta\Delta E)}
where $\wn$ and $\wp$ are the rates for flips down and up, respectively,
and $\Delta E = E_+ - E_-$ is the energy difference of the spin positions.
We are interested in the parameter range where $\wn$ and $\wp$ are nearly
equal to $\alpha$, or $\beta\Delta E$ is small.  Consider an isolated pair of
domain walls separated by distance $\ell$ as shown in
\fig\domwalls{An isolated pair of domain walls separated by a distance $\ell$.
The domain wall on the left can move via a flip of spin A or spin B.  The
difference in the up and down flip rates gives rise to a drift velocity.}.
The domain wall on the left can move through either a spin A flip up
or spin B flip down.  If we assume that the $\wp$ in the neighborhood of
the domain wall are equal to $w(A+)$, and the $\wn$ are equal to $w(B-)$,
then the motion of the wall will be a random walk superimposed over a
slight drift with velocity
\eqn\zag{\eqalign{v_d&=a\,\big[\wn -\wp\big]\cr
&= a\wn\left[1-e^{-\beta\Delta E}\right] \cr
&= a\alpha\beta\Delta E+O\left((\beta\Delta E)^2\right).}}
The energy difference
$\D E=J[U(\ell+a)-U(\ell)]\simeq aJ(dU/d\ell)$ if $\ell\gg a$.  If the
domain wall position is labeled by $x$ (so $d\ell/dx=-1$), then
the drift velocity equation is
\eqn\zah{{dx\over dt}=-a^2\alpha\beta J{\dee U\over\dee x}.}  This can be
generalized to systems of multiple domain walls by considering $U(x-y)$ to
be a pairwise interaction energy which carries with it the appropriate
sign for attractive and repulsive interactions.  By comparison to our original
Langevin equations we identify the kinetic coefficient $\Gamma=a^2\alpha\beta
J$.

Thus far we have neglected the possibility of domain wall pair creation.
The energy for pair creation at distances of order $a$ is small,
and even at low temperatures will occur frequently.  However, the energy
required to create a pair separated at a macroscopic distance $\ell^\p$
is quite large relative to the energy required to move a domain wall
a distance $a$ in the presence of another domain wall at $\ell$.  That
is, for large $\beta$ we can satisfy simultaneously $\beta JU(\ell^\p)\gg 1$
and $a\beta J(dU/d\ell )\ll 1$ for finite $U,dU/d\ell$.  The next
question to address is which process, the random walk or the deterministic
drift, dominates the dynamics at late times.

The characteristic length of a random walk at time $t$ is
\eqn\zai{\eqalign{\ell_{rw}=&(\hbox{number of steps}\times a)^{1\over 2}\cr
=&(a\alpha t)^{1\over 2}.\cr}}
Since $\alpha\propto\Gamma T$ then
\eqn\rwlength{\ell_{rw}\sim T^{1\over 2}t^{1\over 2}.}
There is also a length scale determined by the drift velocity which grows
with time as
\eqn\dlength{\ell_d\sim t^{1/(1+\s)},}
which is found from the equations of motion (see section 3).  The time
dependence of these
length scales determines which process controls the dynamics.  For
$\s>1$ we find $\ell_{rw}>\ell_d$ for large $t$, so a pair of charges can
escape annihilation via a random walk.  This is the dynamical picture of
the disordered phase, as was found in the nearest neighbor Ising
model \BrayIsing.  For $\s=1$ and $T<T_c$ we find $\ell_d>\ell_{rw}$, which
means that a pair of charges can no longer escape annihilation.  For
$\s<1$ also the drift dominates the dynamics at low temperatures.  While this
argument would suggest that this is true for all $T$, it ignores higher order
screening effects which renormalize $J$, causing the random walk effects to
dominate
above the critical point.  When the drift does dominate, the presence of the
random walk should cause at most a finite renormalization the kinetic
coefficient $\Gamma$.  To summarize, the Glauber model of dynamics is
equivalent to the Langevin equations without noise for $\s\le 1$ and $T<T_c$,
and otherwise is equivalent to domain walls undergoing random walks.

\subsec{Related Models}

While $\s\le 1$ is the physically interesting range, the model, without noise,
can be extended to values of $\s>1$.  If the late
time dynamics is described by some renormalization group fixed point,
then this fixed point might be qualitatively similar for all $\s$.
For example, we find
in simulations that, to within our accuracy, the length scale given by the
density grows with power law $t^{1/(1+\s)}$ for
both $\s=1$ and $\s=2$.  In section 4 we show that the dynamical equations
simplify in the large $\s$ limit of this model.  From this limit we can
then work back to study the behavior of models with smaller values of
$\s$.

A similar but more simple system than the Ising model is the $Q$-state
Potts model in the limit of $Q\ginf$.  This model can be mapped to an
interacting defect Hamiltonian which has the same power law interactions as
the Ising model, but only between nearest neighbors.  All other pairs
are non-interacting, which makes this system much easier to simulate on the
computer.
The annihilation rules are modified as well, in that
a pair of defects annihilate to leave behind a single defect.
A derivation of the properties of this model is given in Appendix~A.

\newsec{Scaling Arguments and Numerical Results}

\subsec{Initial Conditions and Scaling Functions}

The initial conditions for the dynamical system are drawn from some
distribution.  Measurements of the system, such as the
density $n(t)$, or the two-particle distribution function
$n_2(r,t)$, are defined to be averaged over this distribution. One
could use a thermal distribution corresponding to $T_0$, the pre-quench
temperature of the system.  Instead we use an initial distribution
where charges are placed randomly with some initial density $n_0$, which
for $n_0=(2a)^{-1}$ corresponds to the system being prepared at
$T=\infty$ prior to quenching.  For values of $n_0<(2a)^{-1}$ the
random distribution is no longer representative of a thermal distribution,
but this approach enables us to explore the sensitivity to initial conditions
without the complication of initial correlations.

To write scaling functions for the quantities such as
the density $n(t)$, we consider all the dimensionful parameters in the
model.  The initial density $n_0$ gives a length scale, as does the
system size $L$ for finite systems.    The lattice spacing has been
taken to zero.  There is one other length scale, given by time.
One way to define this length is by the
range over which an isolated pair of charges will annihilate in time $t$.
For a pair of charges separated by some distance $\ell$ the equations
of motion can be written as a single differential equation
\eqn\zaj{{d\ell\over dt}=-2\Gamma\ell^{-\s}}
which has the solution
\eqn\zak{\ell(t)^{1+\s}=\ell(0)^{1+\s}-2(1+\s)\Gamma t.}
By setting $\ell(t)=0$ we see that the time to annihilation as a function
of the initial distance $\ell$ is
\eqn\zal{t={1\over 2(1+\s)\Gamma}\ell^{1+\s}.}
We rescale the time
\eqn\scalet{2(1+\s)\Gamma t\rightarrow t}
so that the length scale associated with time is
\eqn\zam{\ell_t=t^\z}
where
\eqn\zan{\z={1\over 1+\s}}
is introduced for notational convenience.  This length scale given by
$t^\z$, as well as those of $L$ and $n_0^{-1}$ are the only dimensionful
quantities in the system.  Therefore
\eqn\scalPhi{n(t)=n_0\,\Phi(n_0t^\z,Lt^{-\z}).}
Generally it is assumed that the density does not depend on the
system size, in which case we get the stronger scaling law
\eqn\scalf{n(t)=n_0\,f(n_0t^\z).}

To determine which of these scaling functions apply, we first turn to
numerical simulations (for details, see section 6).  By varying
the initial number of charges $N_0$ and system size $L$ such that the
initial density $n_0$ is unchanged, the system size dependence of the model
can be directly probed.  For $\s=2$ these plots superpose, shown in
\fig\dataa{Simulations for $\s=2$, shown on a log-log plot. The system sizes
used are $L=100$, $200$, $400$, and $800$, and the initial density is fixed
at $n_0=1$.  The data shows no system size dependence.   The power law
$n\sim t^{-1/3}$ is plotted as a visual reference, and is in good agreement
with the data.  The error bars for the data are smaller than the points
plotted.},
implying no system size dependence.  For $\s=1$ we find a slight system
size dependence
(\fig\datab{Simulations for $\s=1$.  The same range of system sizes are used
as in \dataa.  The data for smaller $L$ values exhibits slight system size
dependence.  The power law $n\sim t^{-1/2}$ is plotted as a visual reference.
The statistical error bars are smaller than the points plotted.})
wherein the smaller systems drop below the scaling curve
at late $t$.

We also measure the time dependence of the density for $\s=2$,
and find that it is consistent with the
$\s<1$ prediction of $n(t)\sim t^{-\z}$ for large $t$ \BrayRut.
This result and the scaling form of the density \scalf\ have the
corollary that $n(t)$ is independent of $n_0$.  That is, since
$f(x)=Ax^{-1}$ for large $x$, then
\eqn\zao{n(t)\sim n_0{A\over n_0t^\z}=At^{-\z}.}
We can plot the same data shown in \dataa, but rescaled so all the runs have
the same system size, but different initial densities.  In
\fig\datac{The same data for $\s=2$ as shown in \dataa, but rescaled so
that $L=100$ and the initial density $n_0=1$, $2$, $4$, and $8$.  The
curves collapse to a single function, implying the system is independent of
the initial density at late times.}
we see that
the plots converge to the same function asymptotically.  We propose
that the lack of dependence on the initial length scale may be a general
feature of the late time dynamics.  This is suggestive of $T_0$ independence
for initial conditions corresponding to thermal distributions.

\subsec{Renormalization Group Approach}

The dynamics of the system, for $\s=2$ at least, appears to be scale
invariant at late times.  That is, evolving the system from time $t_1$
to $t_2$, where both times are chosen from the late time regime, is the
same as rescaling the system at $t_1$ by a factor of
\eqn\zap{b=\left(t_2\over t_1\right)^\z.}
Stated another way, the time dependent domain wall probability distribution
is invariant under a rescaling of the system that includes the length scale
of time (but not the initial length scale).  The numerical
data for the density is consistent with this presumed scale invariance,
since $n(t)=At^{-\z}$ is preserved under rescaling $n\rightarrow n/b$ and
$t\rightarrow tb^{1/\z}$.

This scale invariance motivates an analogy to a second order critical
point in equilibrium statistical mechanics, where renormalization
group methods are applicable \KW.  The general RG
approach is a two step process.  First one integrates out the short
distance behavior of the system from some cutoff $a$ to $ba$, creating an
effective theory with modified
coupling constants.  Second, one rescales the system by the factor
$b$, restoring the original value of the cutoff.  The behavior of the
coupling constants under this transformation determine the renormalization
flow of the theory.  The analog of integrating out the short distance
behavior is evolving the system forward in time \BrayRG.  The system is
then rescaled back to the original point in time, giving a modified
theory.  When the system reaches the point where it is invariant under
this transformation, it has reached a stable, self-similar state in
which it will remain for $t\ginf$.

To describe the flow of the theory to its fixed point we define a
Callan-Symanzik $\beta$ function \Amit.  First we define the
renormalized coupling as the density at an arbitrary but fixed late time $\t$.
This is the analog of the normalization point.  The dimensionless coupling
constant, which will be invariant under rescaling, is then
\eqn\gR{g_R=(\Gamma_R\t)^\z n(\t).}
We have restored the time constant $\Gamma$ in the problem, since it is
possible that renormalization effects may cause an effective time
dependence in $\Gamma_R$.  This will be discussed at the end of section
4.  For the purposes of the present argument we will assume that $\Gamma$
is a constant and can be absorbed into a rescaling of time.
A late time correlation function of the system can be expressed either
in terms of the random initial state evolved in time,
or from the normalization point where the initial state information has
been lost.  That is, for some correlation function $G(r,t)$ we have
\eqn\zaq{G(r,t,n_0)=G_R(r,t,g_R,\t).}
The value of $G$ is independent of the normalization scale, so
\eqn\zar{\t{\dee\over\dee\t}G\bigg\vert_{r,t,n_0}=0}
which implies a Callan-Symanzik equation
\eqn\CS{\left[\t{\dee\over\dee\t}+\beta(g_R){\dee\over\dee g_R}
\right]_{r,t,n_0}G_R=0}
where
\eqn\betafn{\beta(g_R)=\t{\dee g_R\over\dee\t}\bigg\vert_{r,t,n_0}.}
If $G_R$ has dimensions (length)$^n$, then dimensional analysis gives
\eqn\diman{\left[-n+{\t\over\z}{\dee\over\dee\t}+{t\over\z}{\dee\over\dee t}
+r{\dee\over\dee r}\right]_{g_R}G_R(r,t,g_R,\t)=0.}
Combining \CS\ and \diman\ to eliminate the explicit $\t$ dependence gives
\eqn\zas{\left[n-{t\over\z}{\dee\over\dee t}-r{\dee\over\dee r}+
\beta(g){\dee\over\dee g}\right]G_R=0.}
If $\beta(g_R^*)=0$ for some value of the dimensionless coupling $g_R^*$, then
\eqn\zat{G_R=r^nh(rt^{-\z})}
which is the self-similar fixed point.  Also, for $g_R(\t)=g_R^*$ we find
\eqn\zau{n(t)=g_R^*t^{-\z},}
the asymptotic form of the density predicted in the energy dissipation
arguments.  The flow into this fixed point
for a given set of initial conditions is determined by the $\beta$ function.
We stress that in this formalism the assumption of a zero of $\beta$ is
mathematically completely equivalent to the statement that
$n(t)t^\z\rightarrow$ constant, but it gives a conceptually different approach
to the problem, and from an approximation standpoint, a method for
extrapolating from the early to the late time regime.
In section 4 we will discuss a method for finding the $\beta$ function
in the large $\s$ limit.

\subsec{Anomalous Behavior for $\s=1/2$}

For the case of $\s=1/2$ we find that the expected power law behavior of
$n(t)\sim t^{-\z}$ is not observed.  The simulations, shown in
\fig\datae{Simulations for $\s=1/2$.  The initial density is fixed at $n_0=1$,
and the system size varies from $L=100$ to $800$.  On the vertical axis
is plotted $\ln t^{2/3}n$, which should be a constant in the scaling regime.
The data shows strong system size dependence, and no range over which the
density has the expected $t^{-2/3}$ power law dependence.  The line drawn
represents $n\sim t^{-0.75}$.  The error bars for the $L=800$ data are
smaller than the size of the points plotted up to $\ln t=4$.},
exhibit less than
convincing power law behavior, and the density decays with an exponent of
at least $-\rho=-0.75$.  From the scaling forms \scalPhi\ and
\scalf\ it follows that this implies either a dependence
on $n_0$ at late times, or dependence the system size $L$, or both.
The data indicates a fairly strong $L$ dependence.

These simulations are quite difficult.  We attempted to reproduce
periodic boundary conditions by including interactions wrapped around
the system, up to some long range cutoff.  For $\s=1/2$ the system is
more sensitive to the cutoff than in the previous cases, and requires
inclusion of many more replicas to simulate periodic boundary conditions (for
a discussion of our methods see section 6).  We were unable to average over
as many realizations of the system, and as a consequence the statistical error
bars in the numerical results are appreciable toward the later times.  The
data is too imprecise to determine whether the system is independent of the
initial density, as was shown for $\s=2$ in \dataa.

We now present a heuristic argument for the lack of dependence on the initial
density, which holds even when the system shows $L$ dependence.
Consider a system which has evolved some very short time $\d t$.  Then
\eqn\zav{\eqalign{n(\d t)=&n_0\Phi(n_0\d t^\z,\d t^\z/L)\cr
=&n_0-2n_0^2\d t^\z+O(\d t^{2\z})\cr}}
where $x=n_0t^\z$, $y=t^\z/L$, and the coefficient of the $\d t^\z$ term,
$\dee\Phi(0,0)/\dee x=2$, is found in section 4.  Also in section
4 we show that there can be no $L$ dependence until at least order
$n_0^3$, so $\dee\Phi(0,0)/\dee y$ is zero.
In this short time $\d t$ the system will build up correlations, but primarily
at short distances.  This short distance information is quickly leaving
the system via annihilation.  We assume that although there are long-distance
correlations building up, they, nevertheless, depend on only one length scale,
given by $n(t)$.  This assumption is expressed
in terms of \scalPhi\ by taking $n_0\rightarrow n(\d t)$ and $t\rightarrow
t-\d t$, so that
\eqn\ansatz{n_0\Phi(n_0t^\z,t^\z/L)=n(\d t)\Phi\left(n(\d t)(t-\d t)^\z,
(t-\d t)^\z/L\right).}
If we expand the right hand side of the equation to order $\d t^\z$
($\z<1$) then
\eqn\zaw{n(t)=n_0\Phi-2\d t^\z n_0^2\Phi-2\d t^\z n_0^2x{\dee\Phi\over\dee x}.}
Setting the $O(\d t^\z)$ term to zero gives the differential equation
\eqn\zax{x{\dee\Phi\over\dee x}=-\Phi}
which has the solution
\eqn\zay{\Phi(x,y)\sim g(y)x^{-1}.}
This argument predicts that the late time behavior will exhibit lack of
dependence on $n_0$, even though it may depend on $L$ through $g(y)$.  The
form of the function $g(y)$ is unspecified, and may play a direct role in
the asymptotic time dependence.

The original argument which led to equation \ansatz\ is difficult to
make rigorous.  The result of the calculation can only be true for
asymptotically late times.  The short distance correlations
take some time to leave the system before the long-distance correlations
dominate the dynamics.

We have also simulated the $Q\ginf$ Potts model for values of $\s=1/2$
and $2$, and found that naive result $n\sim t^{-\z}$ is consistent with
the data for both values of $\s$.  The data is shown in
\fig\Potts{Simulations of the $Q\ginf$ Potts model for $\s=1/2$ and $\s=2$.
The lines for both data sets correspond to the $n\sim t^{-\z}$ curves.
For both values of $\s$ the initial density $n_0=1$ and the system sizes
$L=100$, $300$, and $1000$ are used.  The data superposes very well for
the different system sizes.}.
One might expect to see different behavior from the Ising system, since
in the Potts case there is no need to include multiple wrappings
in the interactions.  For periodic boundary conditions the only requirement
is to include the interaction between the first and last charge.

\newsec{Fugacity Expansion}

\seclab\sfugacity
A technique for calculating the density $n(t)$ and other correlation
functions as expansions in powers of the initial density $n_0$
is developed in this section, and this result is used to calculate the
$\beta$ function defined in section 3 to order $g_R^4$, from which we
estimate the fixed point coupling $g_R^*$.

\subsec{Machinery}

We can use the ideas of equilibrium statistical mechanics to calculate
quantities which are averages
over the distribution of initial conditions.  In doing so it is necessary
to use finite systems, although at the end of the calculation the
$L\ginf$ limit may be taken, if it exists.  The canonical ensemble,
with a fixed initial density, is too difficult to work with, so instead
we use the fixed fugacity or grand canonical ensemble.  One can check
afterwards that the fluctuations in the grand canonical ensemble are of
order $1/\sqrt{L}$.  The
average of some quantity is calculated by expanding in powers of the
fugacity $y$. The coefficient of the $y^k$ term is given by the
integral of this quantity over all the initial conditions for the $k$-body
system.  To normalize these averages we use the analog of the grand
canonical partition function
\eqn\zaz{\Xi=\sum_ky^kV_k}
where $V_k$ is just the volume of configuration space for the $k$-body system.
We will work with ordered charges, so this volume
is $V_k=L^k/k!$.  From this it follows that
\eqn\zba{\Xi=e^{yL}.}
The initial number of charges can be found in terms of the fugacity,
since the value of $N_0$ for a $k$-body system is just $k$.
\eqn\zbb{\eqalign{N(0)&=\Xi^{-1}\sum_ky^kk{L^k\over k!}\cr
&=yLe^{-yL}\sum_k{(yL)^{k-1}\over (k-1)!}\cr &=yL}}
Therefore the fugacity is equal to the initial density $n_0$.

The calculations are actually simpler for a non-periodic system.
It is then important to comment on the boundary conditions, that is the
values of the spins at $x=0$ and $x=L$.  The spin degeneracy factor
of two can be ignored, leaving as the possible boundary conditions either the
spins at each end being equal, or being opposite.
These correspond respectively to there being an even or odd
number of charges in the system.  For convenience, we sum over both cases,
corresponding to free boundary conditions on the Ising spins.

We can use the fugacity expansion to calculate the time-dependent number
of charges $N(t)$.  First we define $N_k(x_1,\dots,x_k,t)$ to be
the number of charges that remain at time $t$, given
$k$ charges at $t=0$ with initial positions $x_1,x_2,\dots, x_k$.
For regions of the
configuration space of initial conditions where no annihilation has
occurred by time $t$, $N_k(t)=k$.  For regions where exactly one annihilation
has occurred by time $t$, $N_k(t)=k-2$.  This continues down to
regions where $N_k(t)=0$ or $1$, after which no more annihilation is
possible.  Integrating $N_k(t)$ over the distribution of initial conditions
gives the coefficient of the $y^k$ term in the fugacity expansion, which we
define to be $Q_k(t)$.  That is
\eqn\zbc{Q_k(t)=\int\limits_{0<x_1<\dots<x_k<L}\prod_{i=1}^kdx_i\>N_k(x_1,
\dots,x_k,t).}
Calculating $Q_k(t)$ for the random distribution is then a process of
partitioning the volume in configuration space by the number of charges at
time $t$, and then summing these regions weighted by their respective
charge numbers.

In general the division of configuration space at time $t$ into regions
of $k$, $k-2$, etc. charges requires solving the $k$-body problem given
by our equations of motion.  The two-body problem can be solved for all
$\s$, and was found in the last section and used to rescale time $t$.
We can use this result to calculate $Q_2(t)$ for
general $\s$ (note that $Q_1(t)=L$ for all $t$).  As defined
\eqn\zbd{Q_2(t)=\int\limits_{0<x_1<x_2<L}dx_1dx_2\>2\,\theta(x_2-x_1-t^\z),}
that is, there is a contribution of $2$ from regions of the integral where
$x_2(0)-x_1(0)$ is greater than the annihilation distance given by $t$,
and a contribution of zero from the rest.  The integration variables are
the {\it initial} positions of the particles.  The time dependence is explicit
in the integrand.  Integrating gives
\eqn\zbe{Q_2(t)=L^2-2Lt^\z+t^{2\z}.}

This allows us to calculate $n(t)$ to order $y^2$.  Expanding $\Xi^{-1}=
e^{-yL}$ to order $y$ gives
\eqn\zbf{\eqalign{N( t)&=(1-yL)(0+yL+y^2L^2-2y^2L t^\z+y^2 t^{2\z})+O(y^3)\cr
&=Ly(1-2y t^\z) + O(L^0,y^3).\cr}}
Dividing both sides by $L$ and taking the $L\ginf$ limit (or just
considering $L\gg t^\z$) gives
\eqn\zbg{n(t)=n_0\left[1-2n_0 t^\z\right]+O(n_0^3t^{2\z})}
or, from \scalf,
\eqn\zbh{f(x)=1-2x+O(x^2).}

\subsec{Large $\s$ Calculation}

The higher order terms become quite difficult.  We can solve
the three-body system for $\s=1$ (see appendix B), but in general
some simplification is needed to proceed.  By taking the large $\s$ limit
the equations of motion effectively decouple, and we can solve for the
higher order terms.  As stated earlier, this limit merits consideration
since the value of $\s$ seems to play only a minor role in the nature
of the fixed point which characterizes the late times, at least for
$\s>1$.  For the three body case the equations of motion
can be reduced to two equations by introducing the variables
$r_i= x_{i+1}-x_i$.  In terms of $r_1,r_2$ the equations of motion are
\eqn\reqs{\eqalign{\dot r_1&=-2r_1^{-\s}+r_2^{-\s}+(r_1+r_2)^{-\s}\cr
\dot r_2&=-2r_2^{-\s}+r_1^{-\s}+(r_1+r_2)^{-\s}.\cr}}
Now suppose $r_1<r_2$.  In the large $\s$ limit the equation of
motion for $r_1$ becomes
\eqn\zbi{\dot r_1=-2r_1^{-\s}}
and the charges have decoupled.  More exactly, the closest pair moves
together and annihilates in a time that is infinitely smaller than the time
scales of the rest of the charges.  With these simplified dynamics we are
able to calculate higher order terms.

For the $k=3$ case we divide our configuration space into two regions
corresponding to the order in which the charges annihilate:  $r_1<r_2$ so
$x_1,x_2$ annihilates first, or $r_1>r_2$ so $x_2,x_3$ annihilates first.
The equations of motion are symmetric with respect to $r_1$ and $r_2$,
so we can consider just one of these conditions, say $r_2>r_1$, and double
the resulting calculation.
We have
\eqn\zbj{Q_3( t)=2\int\limits_{_{0<x_1<x_2<x_3<L}}d^3x\>\theta(r_2-r_1)
\left[3\theta(x_2-x_1- t^\z)+\theta( t^\z-x_2+x_1)\right].}
We can rewrite the square bracket piece as $3-2\theta( t^\z-x_2+x_1)$.
To evaluate this integral it is convenient to take the derivative
with respect to $t^\z$, which turns the $\theta$-function into a
$\delta$-function.
\eqn\zbk{\eqalign{{\dee Q_3( t)\over\dee t^\z}&=2\int_0^Ldx_3\int_0^{x_3}dx_2
\int_0^{x_2}dx_1\>\theta (x_3-2x_2+x_1)(-2)\delta( t^\z-x_2-x_1)\cr
&=-2L^2+8L t^\z-8 t^{2\z}}}
Integrating this we get
\eqn\zbl{Q_3( t)={L^3\over 2}-2L^2 t^\z+4L t^{2\z}-{8\over 3} t^{3\z}}
where the constant of integration is given by  the $t=0$ value,
$Q_k=kL^k/k!$.

To calculate the $k=4$ integral we divide the configuration space into
three regions, distinguishable by which pair annihilates first: $(1,2)$,
$(2,3)$, or $(3,4)$. By symmetry the first and last cases give identical
contributions to the integral.  The next step in evaluating the integral
is to take the derivative $\dee Q_4/\dee L$.  The $L$ dependence
of the integral is contained in
the $\theta(L-x_k)$ term implicit in the limits of integration.  The
$L$ derivative replaces this $\theta$-function with a $\delta(L-x_k)$,
against which
we can integrate $x_k$.  The remaining $k-1$ integrals over the $x_i$
are changed to integrals over $r_i$ with the constraints
$\sum_{i=1}^{k-1}r_i<L$ and $r_i>0$.   Then
\eqn\zbm{\eqalign{ {\dee Q_4(t)\over\dee L}=\int\limits_{r_1+r_2+r_3<L} &d^3r
\bigg[2\theta(r_3-r_1)\theta(r_2-r_1)\{2\theta(r_1-t^\z)+2\theta(r_3-t^\z)\}\cr
+\theta(r_3-r_2)&\theta(r_1-r_2)\{2\theta(r_2-t^\z)+2\theta(r_1+r_2+r_3-t^\z)
\}\bigg].\cr}}
By taking the $t^\z$ derivative as before, the integral can be done fairly
straightforwardly, with the result
\eqn\zbn{{\dee^2Q_4(t)\over\dee t^\z\dee L}=-3L^2+14Lt^\z-{58\over 3}t^{2\z}.}
Integrating this we get
\eqn\zbo{Q_4(t)={L^4\over 6}-L^3 t^\z+{7\over 2}L^2 t^{2\z}-{58\over 9}
L t^{3\z}+\hbox{const.~} t^{4\z}}
where again the initial value of $Q_4$ is used to find the constant of
integration.  The unknown function of $t$ is proportional to $t^{4\z}$,
with a proportionality constant which could be calculated by evaluating the
integral without the $L$ derivative.

On the basis of the scaling relation \scalf\
one might think that the only piece of the $y^k$ integral that is of
interest is the $ t^{(k-1)\z}$ term.  In this case we could take
$k-1$ derivatives with respect to $ t^\z$ and then evaluate the
remaining integral for $ t=0$, a considerable simplification.  However,
it turns out that all the pieces from lower order terms, and not just
the $ t^{(k-1)\z}$~piece, feed back into the calculation of higher order terms.
This is a consequence of boundary effects introduced by working with a
non-periodic system.  Writing a more careful scaling form for
$N(t)$ where both $t$ and $L$ are finite we get
\eqn\zbp{N(t)=Lyf(yt^\z)+g(yt^\z)}
where $f$ is the original scaling function, and $g$ some function which
corresponds to our choice of boundary conditions.  Writing $f(x)=\sum_if_ix^i$
and $g(x)=\sum_ig_ix^i$, we find
\eqn\zbq{\eqalign{\sum_k y^k Q_k( t)=&N( t)e^{yL} \cr
=&Ly(1+f_1y t^\z+f_2y^2 t^{2\z}+f_3y^3 t^{3\z})e^{yL}\cr
&+(g_0+g_1yt^\z+g_2y^2 t^{2\z}+g_3y^3 t^{3\z}+g_4y^4 t^{4\z})e^{yL}
+O(y^5 t^{4\z}).\cr}}
The coefficients for $f$ and $g$ can be determined by comparing powers of $y$
on each side of the equation.  In general, extracting the
coefficient $f_{k}$ from the $k+1$-body integral requires
knowing all the $g_i$ for $i\le k$.  To order $y^4$
we find that $\hbox{$g(x)=x^2-(8/3)x^3+O(x^4)$}$ and
\eqn\fofx{f(x)=1-2x+3x^2-{34\over 9}x^3+O(x^4).}
This density expansion is the main result of this calculation.

With a systematic expansion for the scaling function \scalf\ we have
equivalently an expansion for the $\beta$ function defined by \gR\ and
\betafn\ in powers of $g_R$.  Since $g_R(x)=xf(x)$,
\eqn\zbr{\eqalign{\sigma\beta&=x{d\over dx}g_R(x)\cr
&=x-4x^2+9x^3-{136\over 9}x^4+O(x^5).\cr}}
To find $\beta(g_R)$ we invert the series $g_R(x)$, which gives
\eqn\zbs{\sigma\beta(g_R)=g_R-2g_R^2-2g_R^3-{10\over 3}g_R^4+O(g_R^5).}
The fixed point value of $g_R$ if $\beta$ is truncated at the fourth order
is $g_R^*=0.33$. Truncating to third order would give $g_R^*=0.37$, a
ten percent difference.  The value of $g_R^*$ is the amplitude $A$ in the
asymptotic form of the density
\eqn\zbt{n(t)\sim At^{-\z}.}
This number should be universal in that all systems with the same
value of $\s$ (but different $n_0$) will have the
same amplitude.  We suspect only a weak $\s$ dependence of this number for
values of $\s>1$.  The amplitude found from the numerical data for both
$\s=1$ and $\s=2$ is $A=0.31$.

\subsec{$\s\le 1$ Calculation}

While this approach of calculating the large $\s$ terms may give a
description of the fixed point, our real goal is to work with
values of $\s$ which lie in the range of physical interest.  The two-body
solution is known for all values of $\s$.  For the three-body
term the relevant calculation is the time to the first annihilation,
$T(r_1,r_2)$.  In the large $\s$ limit this was just given by
$T=\min(r_1,r_2)^{1+\s}$.  For finite $\s$ the presence of the third
charge will affect the annihilation of the first and second charges, and
always in the direction of slowing down the process.  This slowing down
will be a maximum when $r_1$ and $r_2$ are approximately equal.  In
\fig\ycubed{Curves of constant time to annihilation, $T(r_1,r_2)=t$, in the
$r_1$, $r_2$ plane for $\s=1$ and the large $\s$ limit.  The
area bounded by these curves, the axes, and the line $r_1+r_2=L$
gives $\dee Q_3(t)/\dee L$, as shown in the text.  The contribution to the
area between the $\s=1$ and large $\s$ curves from the asymptotic region is
divergent as $L\ginf$.  This can be shown by integrating $\D r_1(r_2,t)$
out to $r_2=L$.}
curves of constant $T$ are plotted in the plane of initial conditions
$r_1,r_2$.  The curve for the large $\s$ limit is given by vertical and
horizontal lines,
while the $\s=1$, constant $T$ curve lies to the left and below.
For any value of $\s$ the area bounded by the corresponding constant $T$
curve is proportional to $\dee Q_3(t)/\dee L$, as can be seen by writing out
the integral
\eqn\zbt{{\dee Q_3(t)\over\dee L}=-2\int\limits_{r_1+r_2<L}dr_1dr_2\theta
\big(t-T(r_1,r_2)\big).}
Finding $Q_3(t)$ for $\s=1$ is then a matter of finding the area between the
$T(r_1,r_2)=t$ curves for the large $\s$ limit and $\s=1$.

For $\s=1$
an exact solution for $T(r_1,r_2)$ can be found, the details of which
are given in appendix B.  The result, however, gives an area between
the two curves which diverges as $L\ginf$.  This is a general feature
which occurs for all $\s\le 1$, which can be understood by examining
the equations of motion.  Consider a three charge configuration where
one separation distance, say $r_2$, is much larger than the other.  The
equation of motion for the closer pair is then
\eqn\zbu{2(1+\s)\dot r_1=-2r_1^{-\s}\left[1-{r_1^\s\over r_2^{\s}}+O
\left(r_1^{2\s}\over r_2^{2\s}\right)\right]}
where the factor multiplying the left hand side is a consequence of our
rescaling of time in \scalet.  We treat $r_2$ as a constant in the
equation, and
integrate the dynamical variable $r_1$ from its initial value to zero
\eqn\zbv{-{1\over(1+\s)}\int_0^Tdt=\int_{r_1}^0dr\left[r^{\s}+{r^{2\s}
\over r_2^{\s}}+O\left(r^{3\s}\over r_2^{2\s}\right)\right].}
This gives $T(r_1,r_2)$ in the asymptotic region described.  Performing the
integral and inverting to find $r_1(T,r_2)$ gives
\eqn\zbw{r_1(T,r_2)=T^\z-{T\over (1+2\s)r_2^\s}+O\left(T^{\z(1+2\s)}\over
r_2^{2\s}\right).}
The first term is just the large $\s$ solution, so the second term gives the
leading contribution to $\Delta r_1=r_1^{(\infty)}-r_1^{(\s)}$.  Integrating
$\Delta r_1(r_2)$ out to $r_2=L$ gives the area contained in the asymptotic
approach to the constant $T$ line of the large $\s$ limit.  For $\s=1$ this
piece gives  $\log L$, and for smaller values of $\s$ it gives an $L^{-1+\s}$
term.  The significance
of the divergences is that they will not cancel when the fugacity expansion is
summed, as all the other $L$ dependent terms do.  For $\s>1$
the area remains finite as $L\ginf$, and so the calculation for $\s>1$
should result in the same terms as in the case of the large $\s$ limit, but
with modified coefficients.

It is possible that there may be an infinite set of logarithms (for $\s=1$)
which can be summed to restore the intensive behavior of the density.
Such a summation may then be used, as in conventional critical dynamics \HH,
to renormalize the kinetic coefficient $\Gamma$, effectively
making $\Gamma_R$ a time dependent quantity.  While this would imply no system
size dependence and the density scaling form \scalf, the time dependence
of $\Gamma_R$ would give rise to anomalous time dependence for the density,
as can be seen by \gR.  This anomalous time dependence carries with it the
implication that late time dynamics will exhibit dependence on the initial
density.

An alternate possibility is that these divergent terms
are indicating that the asymptotic dynamics truly has system size dependence.
If the system were still independent of the initial density, as suggested
by our heuristic argument in section 3, then this system
size dependence would give rise to anomalous time dependence, as can be seen
by the scaling function \scalPhi.  It is worth noting that if there is
system size dependence, we can no longer expect our calculations, which
are performed with free boundaries, to correspond directly to simulations
with periodic boundary conditions.

It is possible that both of these effects, system size dependence and a
time dependent $\Gamma_R$, occur.  The simulations for $\s=1/2$, as
discussed in section 3, are not decisive on this issue, although they do
seem to indicate at least the former.

By studying a related model we might hope to find more clues for the
significance of the divergences in the fugacity expansion.  The
$Q\ginf$ Potts model provides a contrast which further confuses the
problem.  In the Ising case the divergences were caused by a three body effect
where the annihilation of a close pair, say $(x_1,x_2)$, is slowed by a
distant charge, $x_3$.  In the Potts case the distant charge is still
interacting with the nearest neighbor, $x_2$, but not with the charge at
$x_1$.  This will give exactly half of the divergent effect seen in the
Ising case.  However, the simulations show no system size dependence, to
within our accuracy.

There is a difference between the two models in the higher order divergent
terms.  Presumably the four body term in the Ising case will have divergent
pieces when one of the end charges,
say $x_4$, is distant, and is affecting both of the possible annihilations:
$(x_1,x_2)$ and $(x_2,x_3)$.  In the $Q\ginf$ Potts case, the distant charge
can only affect the annihilation of the pair which contains
the nearest neighbor of the distant charge, in this case $(x_2,x_3)$.
However, a quantitative analysis of this effect at higher orders is
difficult.

\newsec{Truncation Scheme for the Two-Particle Distribution Function}

The fugacity expansion provides an exact scheme for calculating
time dependent quantities in the system via the deterministic equations of
motion.  A simpler scheme can be developed which gives a
qualitative description of the scaling regime, and of other features of the
model.  In this section we will present this method, and discuss the
applicability for some different initial conditions.

The two-particle distribution function for the system, $n_2(r,t)$,
can be used to find a dynamical equation for $n(t)$.  Integrating
the distribution function from $r=0$ to $r=\d r$ gives the density of
charge pairs which are within $\d r$ of each other.  For very small
separations the charge pairs will become isolated from the rest of the
system, and annihilate in a time $\dt=\d r^{1/\z}$.  Therefore the rate
of change of the density is given exactly by
\eqn\deneq{{dn\over dt}=\lim_{\dt\rightarrow 0}-{2\over \dt}\int_0^{\dt^\z}
n_2(r,t)dr.}

The distribution function can be calculated by the fugacity expansion described
in the last section.  The leading order term is the two-body term, which can
be written
\eqn\twofug{n_2(r,t)\propto\> y^2\int\limits_{r(0)<L}dr(0)
\delta(r(t)-r).}
A change of integration variables from $r(0)$ to $r(t)$ will introduce
the Jacobian
\eqn\jacobian{J(r,t)={dr(0)\over dr(t)}\bigg\vert_{r(t)=r}={r^\s\over
(r^{1+\s}+t)^{\s/(1+\s)}}}
which has the limit $J=1$ for $r\ginf$ or $t=0$.  Therefore the
distribution function is
\eqn\twofugii{n_2(r,t)=n_0^2J(r,t)+O(n_0^3).}
Notice that the dynamics produces a `hole' in the two-particle distribution
function at short distances.

This expansion is only useful for low densities or early times.  However,
we can extend the range via a heuristic argument similar to that of
section 3.  For an isolated pair of charges separated by a distance $r$
at time $t+\dt$, the separation at time $t$ is given by
$\hbox{$(r^{1+\s}+\dt)^\z$}$.  Therefore, for small $r$ we expect
the two-particle distribution
functions of the arguments above to be related.  The relation should also
include the Jacobian, for the same reason it enters into the $t=0$ calculation.
Therefore
\eqn\twopt{n_2(r,t+\dt)=J(r,\dt)\>n_2\left((r^{1+\s}+\dt)^\z,t\right)+
\hbox{higher order terms}.}
This equation should be exact in the small $r$ limit, and the
higher order terms are corrections for large $r$.  The contributions from
the higher order terms can be approximated by replacing $n_2(r,t)$ with
$\nb(r,t)=n_2(r,t)/n(t)^2$, so that $\nb(r\ginf,t)=1$. This results in a
truncation scheme for $\nb$ which is correct both for small $r$ and in the
$r\ginf$ limit.  Making this substitution and equating the order $\dt$ terms
in \twopt\ gives the differential equation
\eqn\twoptb{{\dee\nb\over\dee t}=-\left(\s\over 1+\s\right) {\nb\over r^{1+\s}}
+ \left(1\over 1+\s\right){1\over r^\s}{\dee\nb\over\dee r}}
whose general solution is
\eqn\nbar{\nb (r,t)= r^\s g\left((r^{1+\s}+t)^\z\right).}
For large $x$, we must have $g(x)\sim x^{-\s}$ as determined by the $r\ginf$
limit of $\nb$, corresponding to a scaling solution for the distribution
function
\eqn\twoptii{\nb(r,t)\sim J(r,t).}
{}From \nbar\ we see that
this scaling form of $\nb(r,t)$ will also be the solution for large $t$.
This is consistent with the RG picture of an attractive fixed point which
describes the asymptotically late time dynamics for all initial distributions.

The solution for $\nb$ and equation \deneq\ give two relations between the
density and the distribution function.  Using the small $r$ limit of
the distribution function, $\hbox{$\nb(r,t)=r^\s t^{-\z}$}$, we get the
equation
\eqn\den{{dn\over dt}=-2\z n^2t^{\z-1},}
which is consistent with the scaling solution $n\sim t^{-\z}$.  The amplitude
is not correct, but the argument captures the qualitative features at least.
Note that no system size dependence can appear in this approximation.
For uncorrelated initial conditions the scaling solution for $\nb$ is valid
at $t=0$.  Further qualitatively correct results may be obtained if we assume
\twoptii\ holds for all $t$.  Then the solution to \den\ is
\eqn\denoft{n(t)={n_0\over 1+2n_0t^\z},}
which exhibits the asymptotic time dependence, and also the lack of $n_0$
dependence, we see in the simulations for $\s\ge 1$.  In fact, under the same
assumptions we can find $g(x)$ for all values of $x$ for correlated initial
conditions.  Setting $t=0$ in \nbar\ gives
\eqn\gii{g(x)=x^{-\s}\nb(x,0)}
from which it follows that
\eqn\solution{\nb (r,t)=J(r,t)\>\nb\left((r^{1+\s}+t)^\z
,0\right).}
Combining this with \deneq\ gives
\eqn\denii{{dn\over dt}=-2\z n^2t^{\z-1}\bar n_2(t^\z,0)}
which can be rewritten as
\eqn\deniii{{dn^{-1}\over dt^\z}=2\nb(t^\z,0).}
Thus the late time behavior of the density is completely determined by the
initial two-particle distribution function in this approximation.
We can test \deniii\ by introducing correlations into the initial
conditions.  In particular, if we generate a system via the nearest
neighbor distribution
\eqn\pofx{P(x)=\cases{{\displaystyle 1\over\displaystyle 2n_0\D} &
$n_0^{-1}(1-\D) < x < n_0^{-1}(1+\D)$\cr & \cr 0 & otherwise\cr}}
then $\nb(r,0)$ will be sharply peaked around $r=n_0^{-1}$, less
sharply peaked around $r=2n_0^{-1}$, and so on out to infinity where
it is equal to $1$.  The first $\hbox{$k=(1+\D)/(2\D)$}$ peaks will
have zeroes between them, which implies $dn^{-1}/dt^\z=0$.
Therefore we expect a plot of $n^{-1}$ versus $t^\z$ to have flat areas
separated by sharp jumps, like a staircase, with the jumps
smoothing to a straight line at late times.  The simulation results for these
initial conditions (shown in
\fig\staircase{Simulations for the correlated initial conditions discussed
in the text, for $\s=2$, $\D = 0.05$, and $N_0=100$.
The data is plotted as $n^{-1}$ versus $t^{1/3}$, which should give a
staircase pattern as discussed in the text.  The first three zeros of the
slope are clearly visible.  The data for the random initial conditions is
plotted for reference.})
verify the
the staircase pattern.  Both the truncation method developed here and the
property of lack of dependence on initial conditions are reinforced by this
result.

\newsec{Simulations}

\seclab\ssimulations
To simulate these systems we simply directly integrated the equations of
motion
\eqn\zbx{\dot x_i=\sum\limits_{j\ne i}(-1)^{i+j}|x_j-x_i|^{-\s}
\hbox{sgn}(x_i-x_j).}
Whenever two charges pass each other they were removed from the system.
We began with some number of charges $N_0$ distributed randomly along a
length $L$.  To reproduce periodic boundary conditions exact
replicas of the charge configuration were made, and added to
the left and the right of the original system.  Then the forces were
calculated on the original charges, the positions updated, and the
replicas replaced with updated copies.  While this emulates a periodic
system, it also adds a long-range cutoff to the interactions.  It is important
to separate the effect of the cutoff from possible system size dependence
effects.

We parametrized the cutoff in the following way. If $k$ replicas are added
to the left and to the right of the original system, there are
$N_e=(2k+1)N(t)$ effective charges in the system.  We imposed a
minimum number of effective charges $\Nmin$, and then determined the
number of replicas needed so that $N_e\ge\Nmin$.  By comparing
simulations with different values of $\Nmin$, we could determine at what
point the results are independent of the cutoff to our desired accuracy.
We chose this method of introducing a cutoff, rather than a more obvious
choice of including interactions out to a certain length, because
otherwise small number effects entered into the simulations at late times.
By keeping the effective number of charges at some minimum level we hoped
to more accurately model the truly periodic system.  The values we used for
$\Nmin$ are given in the table.

The numbers for $N_{\rm min}$ are large enough, particularly for $\s=1/2$, to
significantly slow the simulations.  Some speed can be regained by exploiting
the insensitivity of the system to the time step.  For $\s=1/2$ and $n_0=1$
we used an initial time step of $\D t=10^{-3}$, which we stepped up to
$\D t=1$ over the time interval $t=(0,20)$ (for other values of $\s$ see
table 1).  Although these
values for the time step seem large, the results of the simulation are
quite insensitive to the size.  Simulations performed with ten times this
step size showed no appreciable change.  This is somewhat expected, since the
primary consequence of a large time step is to cause the annihilating pairs
to stay around longer, but the force exerted on the system by a very
close pair of charges is nearly zero.  It is probable that an even larger
time step than the one used here would be adequate.

We used the
second order Runge-Kutta integration technique, which involves initially
taking a half-step, reevaluating the forces at this midpoint, then going
back and taking
a full step with the modified forces.  To use this method it is necessary
to check for annihilation at the midpoint, or else the forces calculated
at the midpoint for a pair which has just passed each other will be quite
inaccurate.  It seems
likely that using the Euler integration method instead of Runge-Kutta would
give the same results.

The third parameter in the simulations is the number of runs over which
the quantities are averaged.  That is, since the quantities are supposed
to be averaged over the distribution of initial conditions, we average over
multiple runs.  To determine the number of runs $N_r$ over which to
average, we calculated the standard deviation of $n(t)$ via the central
limit theorem.  For $\s=1/2$ we used $N_r=300$, for which the one-sigma error
bars were smaller than the point size of the plots for much of the scaling
regime.

These simulations were performed on a DECstation 5000.  Our goal was to
use as simple an algorithm as possible, and to keep simulations at the level of
a workstation problem.  There are ways in which our techniques could
be expanded or improved upon, for example by using a controlled
time step which maximally exploits the insensitivity of the system, or of
course, by using faster computers.
It would be preferable to push the $\s=1/2$ system to larger values of $N_0$,
but this was where we were reaching our limits.

\newsec{Discussion}

The first of the principal results of this paper is the appearance of system
size dependence in simulations for $\s=1/2$.  We believe that this
result also holds for all $0<\s\le 1$ on the basis of our fugacity expansion,
which demonstrates anomalous $L$ dependence in the expansion coefficients
for this range.  The data for $\s=1$ exhibits only slight $L$ dependence.
However, the effect may be very small at the marginal value.
This result could bear a more thorough scrutiny via simulations, particularly
if data for larger system sizes can be generated.

The other principal result is the possibly more general dynamical feature of
lack of dependence on the initial length scale.  From a renormalization group
perspective this is an intuitive result:  once the system has reached
the fixed point, and is thus evolving in time via a rescaling of the
domain length only, then the information about the path to the fixed point
is lost.  For the distributions we used, the information lost is
the initial density and correlations.  A variety of distributions may flow
to this same
fixed point, which can then be interpreted as a loss of information about the
initial distribution itself, and not just its length scale.  This is
suggestive of a dynamical feature of lack of dependence on pre-quench
temperature, if thermal distributions flow to the same fixed point.  The
RG picture is complicated by the presence of system size dependence, since
the notion of a scale invariant fixed point will need modification. In the
simulations we find a convincing lack of $n_0$ dependence for $\s=1,2$, and a
possible lack of $n_0$ dependence for $\s=1/2$.    We
also find, via an heuristic argument \ansatz, that the presence of
$L$ dependence should not affect the lack of dependence on the initial
length scale.

Our theoretical approaches include the fugacity expansion and the truncation
method of section 5.  The latter approach is useful in providing a qualitative
description of the dynamics, and includes quite naturally the dynamics at
late times.  In the former case,  the original
intent was to find results for the infinite system while calculating
with finite $L$.  We can find all the expansion coefficients for $\s=0$,
and thus the exact answer for $n(t)$ (see appendix C), but we find that
the density
decays exponentially.  This indicates that $\s=0$ is singular in some
sense, and that trying to expand about this solution is not likely to be
fruitful.  We can calculate the expansion coefficients to
order $y^4$ by taking the large $\s$ limit.  This calculation can be carried
to higher orders, the problem becoming an exercise in bookkeeping.
When we attempt to extrapolate the large $\s$ result
to lower values of $\s$ we find divergences appearing in our expansion
coefficients.  This is a general feature for $\s\le 1$, the physical range
of interest.  These divergences are of the form of $L$ dependent expansion
coefficients, whose presence may just be indicating system size dependence
in the density $n(t)$.

Our expression for the density in the large $\s$ limit can be used to
calculate a Callan-Symanzik $\beta$ function to order $g_R^4$.  To this
order the $\beta$ function has a zero with the value $g_R^*=0.33$ which
predicts the scaling function \scalf\ has the asymptotic form
\eqn\zby{f(x)\sim 0.33x^{-1}.}
This $\beta$ function approach, while useful in giving a qualitative
description of the fixed point, may not give a robust value for
the coefficient of the asymptotic region.  If we take the same $f(x)$
used to derive the $\beta$ function and use the method of Pad\'e approximants,
we find for large $x$
\eqn\zbz{f(x)\sim 0.15x^{-1}.}
Unfortunately, we have not been able to find any analog of the $\epsilon$
expansion  of the equilibrium critical behavior, which would allow a
systematic truncation of the series for the $\beta$ function.

The one-dimensional system with long-ranged interactions appears to have
complicated behavior for $\s\le 1$.  These interactions appear
to be relevant in higher dimensional systems as well \BrayRut, suggesting
that system size dependence may be a more general feature of long-ranged
interacting systems.  Simulations of these systems in higher
dimensions, although difficult, could yield interesting results.  More can
be done with the
one-dimensional simulations in the way of measuring correlation functions
as well.  A theoretical approach which
treats the system size dependence in a controlled way, perhaps some
modification of our density expansion, would be a possible next step in trying
to understand these systems.

\bigbreak\bigskip\bigskip{\centerline{\bf Acknowledgements}}\nobreak
We thank A. Bray and A. Rutenberg for a useful correspondence.  This work
was supported by NSF grant PHY 91-16964.

\appendix{A}{$Q$ State Potts Model}

We can generalize the Ising model Hamiltonian \isingham, which is the
$Q=2$ Potts model, to the case of general $Q$.  At each lattice site there
is a variable $s_i$, which can be in one of $Q$ states.  The Hamiltonian is
defined by
\eqn\zca{H=-\sum\limits_{i<j}V(x_i-x_j)J(s_i,s_j)}
where $V(r)$ is the same as before, and
\eqn\zcb{J(s_i,s_j)=1-\delta_{s_is_j}}
We can rewrite the Hamiltonian as
\eqn\zcc{\eqalign{H=\sum\limits_{i<j}&U(x_i-x_j)\left[J(s_{i+1},s_{j+1})
+J(s_i,s_j)-J(s_{i+1},s_j)-J(s_i,s_{j+1})\right]\cr &+ \hbox{surface terms}.
\cr}}
where the function $U(r)$ is defined by \defu\ \Cardy.
Notice the expression in the square brackets is zero for $s_i=s_{i+1}$
or $s_j=s_{j+1}$.  Therefore this term only contributes to the energy when
$x_i$ and $x_j$ are both locations of defects.  A defect can
be labeled by $(\alpha,\beta)$, meaning it is the boundary between a
region of state $\alpha$ and a region of state $\beta$.  The interaction
for a pair of defects of type $(\alpha,\beta)$ and $(\gamma,\delta)$
separated by a distance $r$ is then
\eqn\zcd{H_{\rm pair}=U(r)\left[\delta_{\alpha\gamma}+\delta_{\beta\delta}
-\delta_{\alpha\delta}-\delta_{\beta\gamma}\right].}

Consider a nearest neighbor pair of defects.  This implies $\beta=\gamma$,
and we assume $\alpha\ne\delta$.  The interaction energy is then
\eqn\zce{H_{\rm pair}=-U(r).}
Also note that if all four states $\alpha,\beta,\gamma,\delta$ are distinct
then there is no interaction between the defects.  Now we take the
$Q\ginf$ limit of this model.  Every domain in the system will find a
unique state, and so all defect pairs will have $\alpha$, $\beta$, $\gamma$,
$\delta$ not equal, with the exception of nearest neighbors.  These rules
allow us to drop the designation of the states, and simply consider the
model to be one where only nearest neighbors interact.
The defect Hamiltonian can be written as
\eqn\zcf{H=-\sum_{i}U(x_i-x_{i+1}).}
We introduce the same equations of motion as before, with the consequence
that now only the
nearest neighbor on either side is included in calculating the force.
There is another modification to the dynamics.  When a nearest neighbor
pair annihilate, a single defect remains.  That is,
\eqn\zcg{(\alpha,\beta)+(\beta,\gamma)\rightarrow (\alpha,\gamma).}

\appendix{B}{$3$-Body Problem for $\s=1$}

We start by
taking the $L$ derivative of $Q_3(t)$, leading to the calculation in the
$r_1,r_2$ plane as shown in \ycubed.  We want to solve for the function which
gives the time to first annihilation, $T(r_1,r_2)$.  From this we can
find $Q_3(t)$ via
\eqn\zcl{{\dee Q_3(t)\over\dee L}=-2\int\limits_{r_1+r_2<L}dr_1dr_2\>
\theta\big(t-T(r_1,r_2)\big).}
The time to annihilation has the scaling form
\eqn\zcm{T(r_1,r_2)=r_2^2f\left(r_1\over r_2\right).}
If $r_1$ and $r_2$ are evolved by a time $\d t$, then $T$ will change by
$-\d t$.  That is, for $r=r_1/r_2$
\eqn\zcn{T-\d t=(r_2+\dot r_2\d t)^2f(r+\dot r\d t),}
so to order $\d t$ we get the equation
\eqn\ftodt{2r_2\dot r_2f(r)+r_2^2\dot rf^\p(r)=-1.}
The three-charge equations of motion given by \reqs\ and the rescaling of
time \scalet\ give
\eqn\zco{\eqalign{{1\over 4}r_2\dot r_2&=-2+{r_2\over r_1}+{r_2\over
r_1+r_2}\cr
&=-2+{1\over r}+{1\over r+1}\cr}}
and
\eqn\zcp{\eqalign{{1\over 4}r_2^2\dot r&=r_2\dot r_1-r_1\dot r_2\cr
&=-{2\over r}+2r+{1\over r+1}-{r\over r+1}.\cr}}
Therefore $r_2$ can be eliminated from equation \ftodt, giving the
differential equation for $f(r)$
\eqn\zcq{{2(1-2r^2)\over(r-1)(2r^2+3r+2)}f(r)+f^\p(r)={-4r(r+1)\over(r-1)
(2r^2+3r+2)}.}
Notice that the coefficients are singular at $r=1$.

This equation can be integrated in closed form, which is somewhat
surprising, with the result
\eqn\zcs{f(r)={4\over 3}(1+r+r^2)+C(r-1)^{2/7}(2r^2+3r+2)^{6/7}}
where $C$ is a constant of integration.  To determine $C$ we consider the
large $r$ limit of $f(r)$.  If $r_1\gg r_2$ then the time to annihilation
is given by the separation $r_2$ only, and is $T=r_2^2$.
This implies for large $r$, $f(r)=1$.
If we choose as our integration constant $C=-2^{8/7}/3$, then the $r^2$ and
$r$ parts of $f(r)$ have coefficients of zero.
Therefore, the exact solution is
\eqn\zcu{f(r)={4\over 3}(1+r+r^2)-{2^{8/7}\over 3}(r-1)^{2/7}(2r^2+3r+2)^{6/7}.
}
When we plot $T(r_1,r_2)$ on the $r_1,r_2$ plane (see \ycubed) we see a
cusp at $r_1=r_2$.  The appearance of the exponent $1/7$ is curious.   When
we use this solution to calculate the area between the $\s=1$ and the large
$\s$ curves, we find this area is divergent, as mentioned in the text.

\appendix{C}{$\s=0$ Solution}

For the $\s\rightarrow 0$ limit of the model we have the equations of
motion
\eqn\zch{\eqalign{\dot x_i&=-{\dee\over\dee x_i}\sum\limits_{j<k}(-1)^{j+k}
|x_j-x_k|\cr &=\sum_{k\ne i}(-1)^{k}\hbox{sign}(i-k)\cr}.}
The force on a given charge does not depend on the position of its neighbors,
only on the global excess of charge on either side. It is necessary to
consider only systems with even numbers of charges, since a system with an odd
number of charges will have all forces equal to zero.  For an even charge
system the charges will be attracted in isolated pairs.
That is, the leftmost charge, call
it positive, will see a net negative charge to the right.  The second charge
from the left will be negative and see only a net positive charge to the
left.  This pair will then move toward each other an annihilate independent
of the rest of the system.

The time dependent density in this model is entirely determined by the
probability distribution for the location of the nearest neighbors.  For the
random initial conditions we used, this is a Poisson distribution
\eqn\zci{P(x)dx=n_0e^{-n_0x}dx}
where $P(x)dx$ is the probability of the nearest neighbor being located
between $x$ and $x+dx$.  At a given time $t$ all the paired charges which
are located within a range $\D x(t)$ will have annihilated.  For the
rescaled time given by \scalet\
\eqn\zcj{\D x(t)=t.}
The fraction of initial charges which remain at time $t$ is then
\eqn\zck{\eqalign{{n(t)\over n_0}&=1-\int_0^tdxn_0e^{n_0 x}\cr
&=e^{-n_0 t}\cr}}
so the density scaling function \scalf\ is
\eqn\zck{f(x)=e^{-x}.}
This result can be found also by using the fugacity expansion for $\s=0$.

\listrefs

\listfigs

\vbox to 1in{\vfil}
\def\row#1,#2,#3,#4,#5 {\hbox{
\vbox to .3in{\vfil\hbox to .7in{\hfil#1\hfil}\vfil}
\vrule\vbox to .3in{\vfil\hbox to .7in{\hfil#2\hfil}\vfil}
\vrule\vbox to .3in{\vfil\hbox to .7in{\hfil#3\hfil}\vfil}
\vrule\vbox to .3in{\vfil\hbox to .7in{\hfil#4\hfil}\vfil}
\vrule\vbox to .3in{\vfil\hbox to .7in{\hfil#5\hfil}\vfil}
\vrule}}
\centerline{\hbox{\vrule
\vbox{\hrule\row ,$N_r$,$N_{\rm min}$,$\Delta t$,$t_0$ \hrule
\row $\sigma=1/2$,300,400,1,20 \hrule
\row $\sigma=1$,1000,30,.1,2 \hrule
\row $\sigma=2$,1000,20,.03,.6 \hrule
}}}
\bigskip
\noindent Table 1.  Values of the simulation parameters used.  For all
simulations $n_0=1$ and $\hbox{$N_0=100, 200, 400, 800$}$.  The time step
is ramped from $10^{-3}\D t$ to $\Delta t$ in the range $\hbox{$0<t<t_0$}$,
and is equal to $\D t$ for $t>t_0$.

\bye